 \documentclass[final,5p,times,twocolumn,authoryear,numbers, square, sort&compress]{elsarticle}

\usepackage{epsfig}

\usepackage{amssymb}
\usepackage{lipsum}
\usepackage{amsmath}
\usepackage{amsthm}
\usepackage{graphicx,framed} 
\usepackage{dcolumn} 
\usepackage{bm} 
\usepackage{braket}
\usepackage{dsfont}
\usepackage{color}
\usepackage{amsthm}
\usepackage{flushend}

\usepackage[normalem]{ulem} 
\usepackage{qcircuit}       
\usepackage[hidelinks]{hyperref}
\usepackage{amsthm}

\usepackage{comment}
\usepackage{float}
\setcitestyle{numbers}

\makeatletter
\def\ps@pprintTitle{%
     \let\@oddhead\@empty
     \let\@evenhead\@empty
     \def\@oddfoot{\reset@font\hfil}%
     \let\@evenfoot\@oddfoot
}
\makeatother

\usepackage{caption}
\usepackage{stfloats}

\journal{}

\begin{document}

\begin{frontmatter}



\title{Concurrence-Driven Path Entanglement in Phase-Modified Interferometry}


\author{H. O. Cildiroglu}
\affiliation{organization={Boston University },
            addressline={Physics Department}, 
            city={Boston},
            postcode={02100}, 
            state={MA},
            country={USA}}
\affiliation{organization={Ankara University },
            addressline={Department of Physics Engineering}, 
            postcode={06100}, 
            state={Ankara},
            country={Türkiye}}

\begin{abstract}
In this study, a novel experimental setup analogous to joint spin/polarization measurement experiments is proposed by establishing a direct relationship between path (momentum) entanglement and concurrence. The results demonstrate that joint-detection probabilities can be governed not only by phase shifts but also by concurrence, which arises from the angle between the motion direction of particles from the same source and the Beam Splitter (BS) axis. This approach aims to set a new standard in entanglement measurement by integrating path entanglement within a concurrence-based framework. Here, we first examine phase-retarder-modified Mach-Zehnder (MZ) configurations within single-quanton systems, subsequently extending this approach to two-quanton systems to establish a connection between spatial correlations and concurrence. Last, by analyzing joint-detection probabilities across various BS configurations, we evaluate the potential of these setups as analogs for spin/polarization measurement experiments.

\end{abstract}




\end{frontmatter}



\section{Introduction}

In classical mechanics, the state of a physical system is the precise information of the position and momentum of all degrees of freedom constituting the system at any given time. However, in the quantum paradigm, the inability to precisely determine such dynamic parameters simultaneously leads to a radical change of the state concept, now characterized by wave functions. Within this framework, physical quantities are described by probability distributions, emphasizing the novel significance ascribed to the wave function. Composite systems are formed by combining two or more independent subsystems, which allows for quantum correlations. When the state of the composite system can be represented as a single tensor product of discrete and normalized states, it is classified as a product state. Conversely, if no such representation exists, the system is identified as a pure entangled state, capturing the intrinsic quantum correlations between its subsystems.

Entanglement measures \cite{Wootters1998,Vedral1997,Vidal2002,Plenio2007,Uhlmann2000,Rungta2001,Horodecki2009} are crucial in emerging quantum technologies, impacting fields such as quantum computing \cite{Raussendorf2003, Walther2005}, ion trapping \cite{Millman2007,FossFeig2022}, and precision sensor devices \cite{Degen2017, Brady2022} that surpass classical limitations. For two-quanton systems, the simplest and most compact method to determine the entanglement measure is the concurrence \cite{Wootters1998}. The concurrence numerically expresses the strength of entanglement between two quantons, ranging from 0 to 1, and provides a practical and concise indication of the correlation between the subparts. Accordingly, when $C=0$, the system represents product states in which the parts are independent, whereas $C=1$ indicates the existence of maximally entangled quantum states. These experiments are often conducted by examining correlations in light polarizations within complex setups using Mach-Zehnder (MZ) type interferometers \cite{Zeilinger1981,Horne1989}, or by analyzing spin correlations of moving particle in Bell-type experiments \cite{Bell1964,Bohm1957,Tunalioglu2023,Cildiroglu2024}. Considering this framework, path (momentum) entanglement has also consistently emerged as a theoretically intriguing option \cite{Einstein1935,Fickler2014,Krenn2017,Kysela2020,Runeson2023,Yang2024}. However, the fact that measurement probabilities for particles in maximally entangled states can only be controlled by phase shifts limits experimental flexibility and makes achieving precision challenging in the current theory.

In this study, a direct relationship between path entanglement and concurrence is established, and a novel experimental setup is proposed to effectively utilize path entanglement in entanglement measurement experiments. The results demonstrate that the measure of the entanglement can be controlled not only through phase shifts but also by the angles between the direction of motion of particles produced from the same source with the Beam Splitter (BS) axis, and thus by concurrence. This approach provides substantial experimental flexibility and enables the construction of analogous setups for polarization/spin experiments. Consequently, our study aims to establish a new standard in entanglement measurement experiments by integrating path entanglement with a concurrence-based framework. 

This letter is organised as follows: In the second section, we focus on the use of MZ-type setups, modified by the phase retarder for one-quanton systems, considering the state of the system in new path-associated bases. In the third section, we first generalize this framework to two-quanton systems, establishing a connection between concurrence through the angles made with the BS axis. This formulation incorporates spatial correlations derived from momentum conservation in all directions within a comprehensive theoretical model. We then analyse in detail the joint-detection probabilities obtained depending on C and P in two possible BS scenarios. Last, we demonstrate that the probabilities derived in the proposed setups are equivalent to joint spin measurement probabilities.

\begin{figure*}
    \centering
    \includegraphics[width=0.98\textwidth]{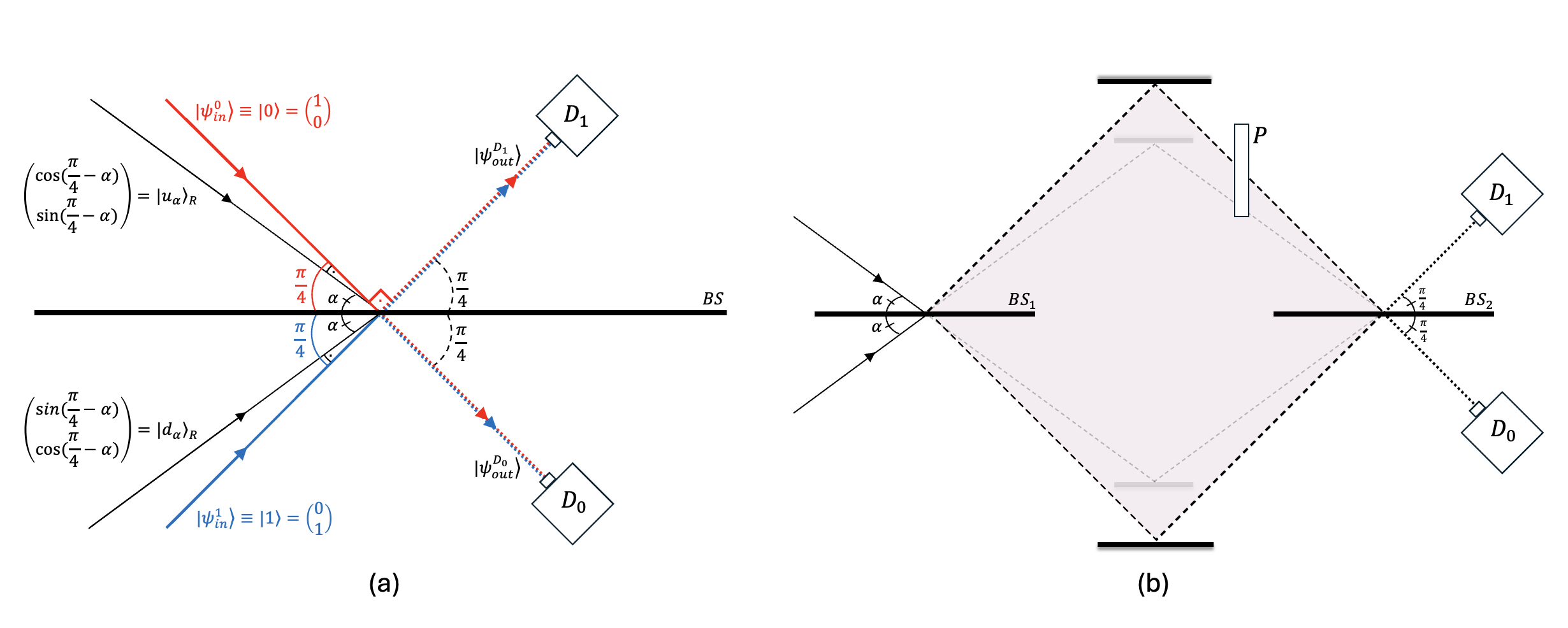}
    \caption{(a) Single quanton system passing through a two-port, lossless (50:50) BS in arbitrary $\alpha$ directions. (b) Schematic representation of incoming quantons generated in arbitrary $\alpha$ directions in MZ-type interferometers modified with phase retarder.}
        \label{fig1}
\end{figure*}

\section{Single quanton systems}

Let us consider a single-quanton system generated from a source $S$ in a special direction ($\alpha=\pi/4$) and reaching a lossless (50:50) BS with two input (red and blue paths in Fig. ~\ref{fig1}.a) and two output ports ($ D_0 $ and $ D_1 $) . To describe the state of the incoming quantons one can use the basis vectors  corresponding to input ports $ (1, 0)^T \equiv |0\rangle $ and $ (0, 1)^T \equiv |1\rangle$. Similarly, for the output states the basis vectors $ (1, 0)^T $ and $ (0, 1)^T $ can be chosen to represent the states through the output ports . The BS applies a linear transformation that converts the incoming state $ |\psi_i\rangle $ into an outgoing state $|\psi_{out}\rangle $. Assuming the (real) reflection amplitudes $ r_i = R_i $ and (purely imaginary) transmission amplitudes $ t_i = iT_i $ (where $ r_0 $ and $ t_0 $ correspond to quantons entering through port 0, and $ r_1 $ and $ t_1 $ correspond to quantons entering through port 1), the conditions $ |r_0| = |r_1| = |t_0| = |t_1| $, and $ |r_i|^2 + |t_i|^2 = 1 $ hold along with $ r_0^* t_1 + t_0^* r_1 = 0 $ \cite{Zeilinger1981}. Thus, the beam splitter can be represented as an operator in the form of a unitary matrix:

\begin{equation}
\label{eq:1}
BS=\left(\begin{array}{cc} r_{0} & t_{1}\\
t_{0} & r_{1} \end{array}\right)=\frac{1}{\sqrt{2}}\left(\begin{array}{cc} 1 & i\\
i & 1 \end{array}\right)
\end{equation}

\noindent Hence, the output state for a single incoming quanton through port $i$ can be expressed as, $\left|\psi_{in}^{0(1)}\right\rangle \rightarrow \left|\psi_{out}\right\rangle = \frac{1}{\sqrt{2}}\left[\left|D_{0(1)}\right\rangle + i\left|D_{1(0)}\right\rangle \right]$. In this case, the detection probabilities for each detector are $\frac{1}{2}$.

For quantons generated from S in arbitrary directions $\alpha$ — where the incoming quantons deviate from the red and blue paths depicted in in Fig. ~\ref{fig1}.a — generalization of this formulation is required. Here, to describe the motion of the particles, while considering a path-entangled setup based on momentum conservation established in the subsequent analysis, wlog, we prefer that the angles made with the BS from both directions are the same. Accordingly, for the quantons approaching the BS from up and down at the angles $\alpha$, we use the notations $|u_\alpha\rangle$ and $|d_\alpha\rangle$, respectively:

\begin{equation}
\label{eq:2}
|u_\alpha\rangle=\begin{pmatrix}
\cos\left(\frac{\pi}{4} - \alpha\right) \\
\sin\left(\frac{\pi}{4} - \alpha\right)
\end{pmatrix} \quad
|d_\beta\rangle=
\begin{pmatrix}
\sin\left(\frac{\pi}{4} - \alpha\right) \\
\cos\left(\frac{\pi}{4} - \alpha\right)
\end{pmatrix}
\end{equation}

\noindent For $\alpha = \pm\pi/4$, particles align with the red and blue paths. When $\alpha = 0$ and $\alpha = \pi/2$, particles are produced horizontally and vertically. Here, the detectors are fixed at angles of $\pi/4$ relative to the BS. In this configuration, the final state for particles approaching from above is given by $|u_\alpha\rangle' = [BS]|u_\alpha\rangle = \frac{1}{\sqrt{2}} \left[ e^{i\left(\frac{\pi}{4} - \alpha\right)}|D_0\rangle + e^{-i\left(\frac{\pi}{4} - \alpha\right)} |D_1\rangle \right]$. The detection probabilities for particles reaching each detector remain $1/2$.

Besides, the use of BS in combination with P in MZ-type setups gives rise to noteworthy effects. In such setups, the phase of the retarder directly controls the quanton detection probabilities, and can be used analogously to spin/polarization measurement angles \cite{Bertlman2003,Hasegawa2003,Sponar2010,Cildiroglu2019,Cildiroglu2024}. Hence, the measurement probabilities along the corresponding angles align with the probabilities of particle detection. In accordance with the existing configuration, wlog, for incoming quantons from the upper arm (See Fig. ~\ref{fig1}.b), the retarder operator $P$ can be expressed as,

\begin{equation}
\label{eq:3}
P=\left(\begin{array}{cc} e^{i\theta} & 0 \\
0 & 1 \end{array}\right)
\end{equation}

\noindent We consider the MZ as a system and focus on the incoming particles with the angles $\alpha$. Moreover, choosing the paths of MZ as $\alpha$ dependent does not affect the detection probabilities. In such a case, for an incoming single quanton through port $i = {u, d}$, its wavefunction $|\psi^{(i)}_{\text{in}}\rangle$ evolves into $|\psi_{\text{out}}\rangle = [\text{BS}_2][P][\text{BS}_1] |\psi^{(i)}_{\text{in}}\rangle$. Accordingly, it can be expressed in a compact representation, where $\alpha' = \left(\frac{\pi}{4} - \alpha\right)$.

\begin{equation}
\label{eq:4}
   \begin{pmatrix}
  |u_\alpha\rangle \\
 |d_\alpha\rangle
\end{pmatrix}
\rightarrow 2ie^{i\frac{\theta}{2}}
\left(\begin{array}{cc} \sin{(\frac{\theta}{2}+\alpha')} & \cos{(\frac{\theta}{2}+\alpha')} \\ \cos{(\frac{\theta}{2}-\alpha')} & -\sin{(\frac{\theta}{2}-\alpha')} \end{array}\right)
\begin{pmatrix}
|D_0\rangle \\
|D_1\rangle
\end{pmatrix}
\end{equation}

\begin{figure*}[h]
    \centering
    \includegraphics[width=\textwidth]{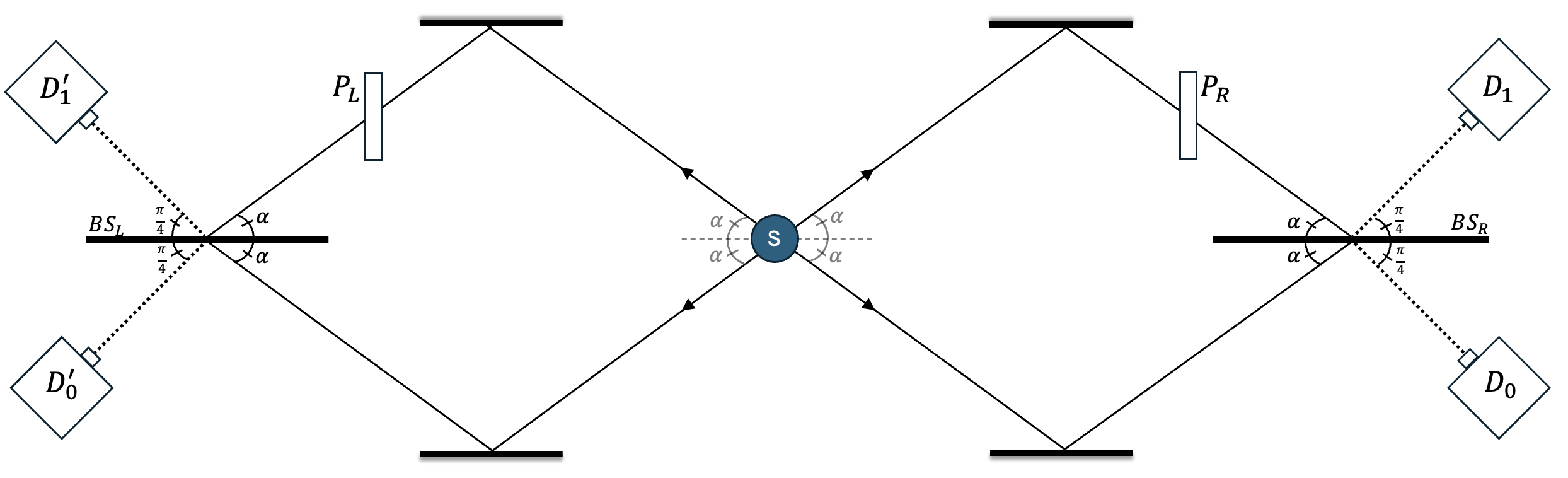}
    \caption{The illustration of the BS-P configurations is proposed for two-quanton systems with spatial correlations generated from a source. Here, quantons enter the system at arbitrary angles, enabling exploration of entanglement across various path orientations.}
    \label{fig2}
\end{figure*}

\noindent Here, $ie^{i\frac{\theta}{2}}$ is a common phase factor and can be eliminated, as it does not contribute to the expectation value expressions.

\section{Two-quanton systems}

Now, we extend the setups proposed in the previous section to two-quanton systems, as this approach is optimal for defining correlated states. A spatial correlation can be established via conservation of momentum for particles generated from the S source, in a process similar to the decay of an atom. Accordingly, in the most general case, a quanton sent upward with an angle $\alpha$ $|{u_{\alpha}}\rangle_R$ on the right side, accompanied by a corresponding particle on the left side moving downward at the same angle $|d_\alpha\rangle_L$. Similarly, a quanton moving downward to the right $|d_\alpha\rangle_R$ is paired with a quanton moving upward to the left $|u_\alpha\rangle_L$. For $0 < \alpha < \pi/2$, such a correlation can be expressed as follows:

\begin{equation}
\label{eq:5}
    |\psi\rangle=|{u_{\alpha}}\rangle_R \otimes |d_\alpha\rangle_L + |d_\alpha\rangle_R \otimes |u_\alpha\rangle_L
\end{equation}

\noindent Here, we note that for the values of $\alpha$ greater than $\pi/2$, the paths overlap, and the symmetry of the problem remains intact. By replacing (\ref{eq:4}) within the equation (\ref{eq:5}), 

\begin{equation}
\label{eq:6}
    |\psi\rangle=\begin{pmatrix}
\cos{\alpha'} \\
\sin{\alpha'}
\end{pmatrix} \otimes \begin{pmatrix}
\sin{\alpha'} \\
\cos{\alpha'}
\end{pmatrix} + \begin{pmatrix}
\sin{\alpha'} \\
\cos{\alpha'}
\end{pmatrix} \otimes \begin{pmatrix}
\cos{\alpha'}\\
\sin{\alpha'}
\end{pmatrix}
\end{equation}

\noindent or equivalently, by considering the z-basis ${{|0\rangle, |1\rangle}}$ and the transformation $\sin{(\frac{\pi}{2}-2\alpha)}=\cos{2\alpha}$, the following expression is obtained:

\begin{equation}
\label{eq:7}
|\psi\rangle= \frac{1}{N(\alpha)}\bigg[ \cos{2\alpha} \left[|00\rangle+|11\rangle\right] + \left[|01\rangle 
+ |10\rangle\right] \bigg]
\end{equation}


\noindent The wave function is normalized by the factor $N(\alpha) = {\sqrt{2[1+\cos^2{2\alpha}]}}$. It is evident that the system can be characterized based on the values of $\alpha$. Accordingly, when $\alpha=0$, a product state $|\psi\rangle=(|0\rangle+|1\rangle)_R \otimes (|1\rangle+|0\rangle)_L/2$ emerges, whereas for $\alpha=\pi/4$, the system reaches the maximally entangled state $|\psi\rangle=(|01\rangle + |10\rangle)/\sqrt{2}$. This result clearly implies that the state and correlations can be controlled via the concurrence.

\begin{figure}[ht]
    \begin{minipage}{\linewidth} 
        \centering
        \includegraphics[width=0.9\linewidth]{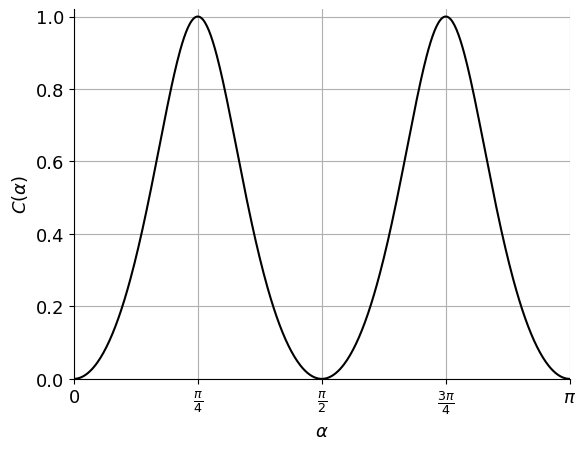}
        \caption{The concurrence as a function of the angle $\alpha$. The function exhibits periodicity for the values of $\alpha>\pi$. For $C=0$, the system is initially in a product state, while for $C=1$, the system is maximally entangled.}
        \label{fig:Concurrence}
    \end{minipage}
\end{figure}

The concurrence (C) is a simple, effective, and preferred measure of entanglement \cite{Wootters1998}. It provides a practical tool for determining entanglement and serves as a powerful approach for analysis. For a general two-quanton physical system with $|\psi\rangle = a|00\rangle + b|01\rangle + c|10\rangle + d|11\rangle$,  the concurrence is defined as $C = 2\left|{ad - bc}\right|$ in its simplest form. Thus, the concurrence of the system given in \eqref{eq:7} is as follows:

\begin{equation}
    \label{eq:8}
    C(\alpha)=\left|\frac{1-\cos^2{2\alpha}}{1+\cos^2{2\alpha}}\right|=\frac{\sin^2{2\alpha}}{1+\cos^2{2\alpha}}
\end{equation}

\noindent The concurrence of this pure state can take values in the range $0 \leq C(\alpha) \leq 1$, where $\alpha = 0 \rightarrow C = 0$ corresponds to a product state, and $\alpha = \frac{\pi}{4} \rightarrow C = 1$ represents a maximally entangled state (see Fig. \ref{fig:Concurrence}). Hence, by considering \eqref{eq:8}, the state \eqref{eq:7} can be rewritten in terms of concurrence in a compact form,

\begin{equation}
    \label{eq:9}
    |\psi\rangle= \frac{\sqrt{1-C}}{2}\left[|00\rangle+|11\rangle\right] + \frac{\sqrt{1+C}}{2}\left[|01\rangle + |10\rangle\right]
\end{equation}

\noindent In this context, the system given in \eqref{eq:9} for  spatially correlated quantons through momentum conservation can be utilized to observe the path-entanglement, to perform entanglement measurements, or implement advanced quantum mechanical applications \cite{Berry1984,Aharonov1984,Allman1992,He1993,Wilkens1994,Bright2015,Cildiroglu2018,Cildiroglu2018Dual,Hashemi2019,Claeys2019,choudhury2019,Jing2020,Cildiroglu2024b,wakamatsu2024,Wang2024}. Therefore, we now concentrate on two fundamental scenarios.

\vspace{\baselineskip}

\noindent \textbf{\textit{P-BS systems:}} In the first scenario (see Fig. \ref{fig2}), the quantons are produced from the S in $\alpha$-angled reverse paths, and before reaching the BSs, they are passed through phase shifters placed in one of the opposing arms. Accordingly, the initial state of the system can be given by \eqref{eq:9}, and the final state of the system $\theta_R \pm \theta_L = \theta_\pm$, as $|\psi_{out}\rangle = (BS_R \otimes BS_L)(P_R \otimes P_L)|\psi\rangle$ can be expressed in the basis of the detectors placed on the left and right sides, with $\theta_R \pm \theta_L = \theta_\pm$,

\begin{eqnarray}
    \label{eq:10}
    |\psi_{out}\rangle= \frac{\sqrt{1-C}}{4}\bigg[ e^{i\theta_+}(|D_0\rangle+i|D_1\rangle)\otimes(|D^{'}_0\rangle+i|D{'}_1\rangle) \nonumber\\
    +e^{-i\theta_+}(i|D_0\rangle+|D_1\rangle)\otimes(i|D^{'}_0\rangle+|D^{'}_1\rangle) \bigg] \nonumber \\
    +\frac{\sqrt{1+C}}{4}\bigg[e^{-i\theta_-}(|D_0\rangle+i|D_1\rangle)\otimes(i|D^{'}_0\rangle+|D{'}_1\rangle) \\
    +e^{-i\theta_-}(i|D_0\rangle+|D_1\rangle)\otimes(|D^{'}_0\rangle+i|D^{'}_1\rangle)\bigg]\nonumber
\end{eqnarray}

\noindent Thus, by eliminating the common phase factor $e^{i(\pi/2)}$, and the joint-detection amplitudes can be found by $A(D_i, D^{'}_j)=\langle D_i D^{'}_j|\psi_{out}\rangle$ where $(i,j=0,1)$,

\begin{eqnarray}
    \label{eq:11}
    A(D_0, D^{'}_0)= \frac{\sqrt{1-C}}{2} \sin{\theta_+}+ \frac{\sqrt{1+C}}{2} \cos{\theta_-} \nonumber\\
    A(D_1, D^{'}_1)= -\frac{\sqrt{1-C}}{2} \sin{\theta_+}+ \frac{\sqrt{1+C}}{2} \cos{\theta_-} \nonumber\\
    A(D_0, D^{'}_1)= \frac{\sqrt{1-C}}{2} \cos{\theta_+}+ \frac{\sqrt{1+C}}{2} \sin{\theta_-} \\
    A(D_1, D^{'}_0)= \frac{\sqrt{1-C}}{2} \cos{\theta_+}- \frac{\sqrt{1+C}}{2} \sin{\theta_-} \nonumber
\end{eqnarray}

\noindent Using the amplitudes, the joint-detection probabilities can be reached: $P(D_i, D^{'}_j)= \left| A(D_i, D^{'}_j) \right|^2$. Although probabilities are easy to calculate, they are in quite complex structure. To simplify this relative complexity, wlog, let us continue by choosing the phase of one of the retarders as $\theta_L=0$. In this case, with $\theta_-=\theta_+=\theta$, the probability expressions $P(\alpha, \theta)$ is reached as,

\begin{eqnarray}
    \label{eq:12}
    P(D_0, D^{'}_0)=\frac{1}{4} \bigg[ 1 + C \cos{2\theta} + \sqrt{1-C^2} \sin{2\theta} \bigg] \nonumber\\
    P(D_1, D^{'}_1)=\frac{1}{4} \bigg[ 1 + C \cos{2\theta} - \sqrt{1-C^2} \sin{2\theta} \bigg]\nonumber\\
    P(D_0, D^{'}_1)=\frac{1}{4} \bigg[ 1 - C \cos{2\theta} + \sqrt{1-C^2} \sin{2\theta} \bigg]\\
    P(D_1, D^{'}_0)=\frac{1}{4} \bigg[ 1 - C \cos{2\theta} - \sqrt{1-C^2} \sin{2\theta} \bigg]\nonumber
\end{eqnarray}

\begin{figure}[H]
    \begin{minipage}{\linewidth}
        \centering
        \includegraphics[width=\linewidth]{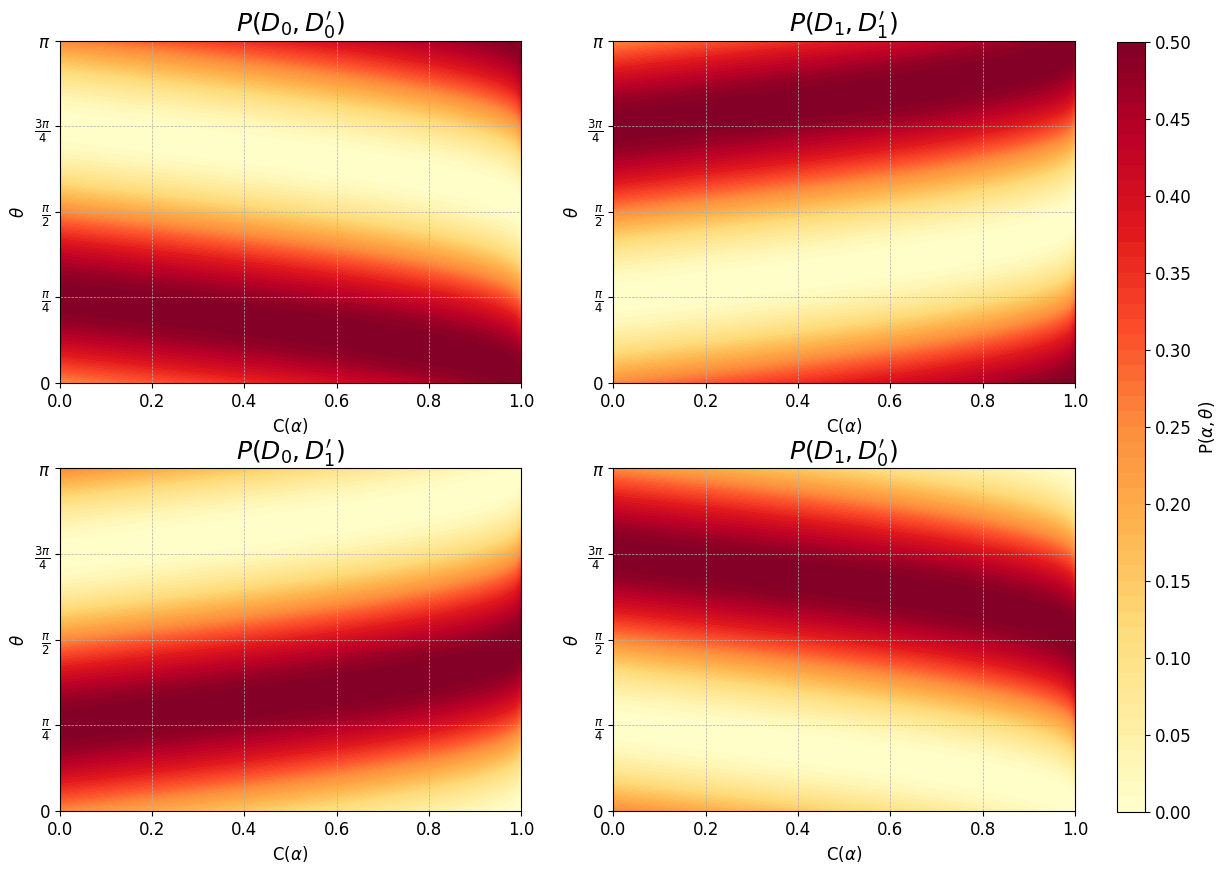}
        \caption{The probabilities in \eqref{eq:12} appear as a function dependent on $\theta$ and $C(\alpha)$, and consequently, the generation angles from the S. Accordingly, a two-dimensional heat map of $P(\alpha, \theta)$ is presented here. The graphs should be evaluated for the same $(\alpha, \theta)$ pairs. In the dark red regions of the graphs $P=1/2$, indicating the presence of maximally entangled states. In the light yellow regions $P=1/4$, indicating that the system is in the product state. \vspace{\baselineskip}}
        \label{fig4}
        \includegraphics[width=\linewidth]{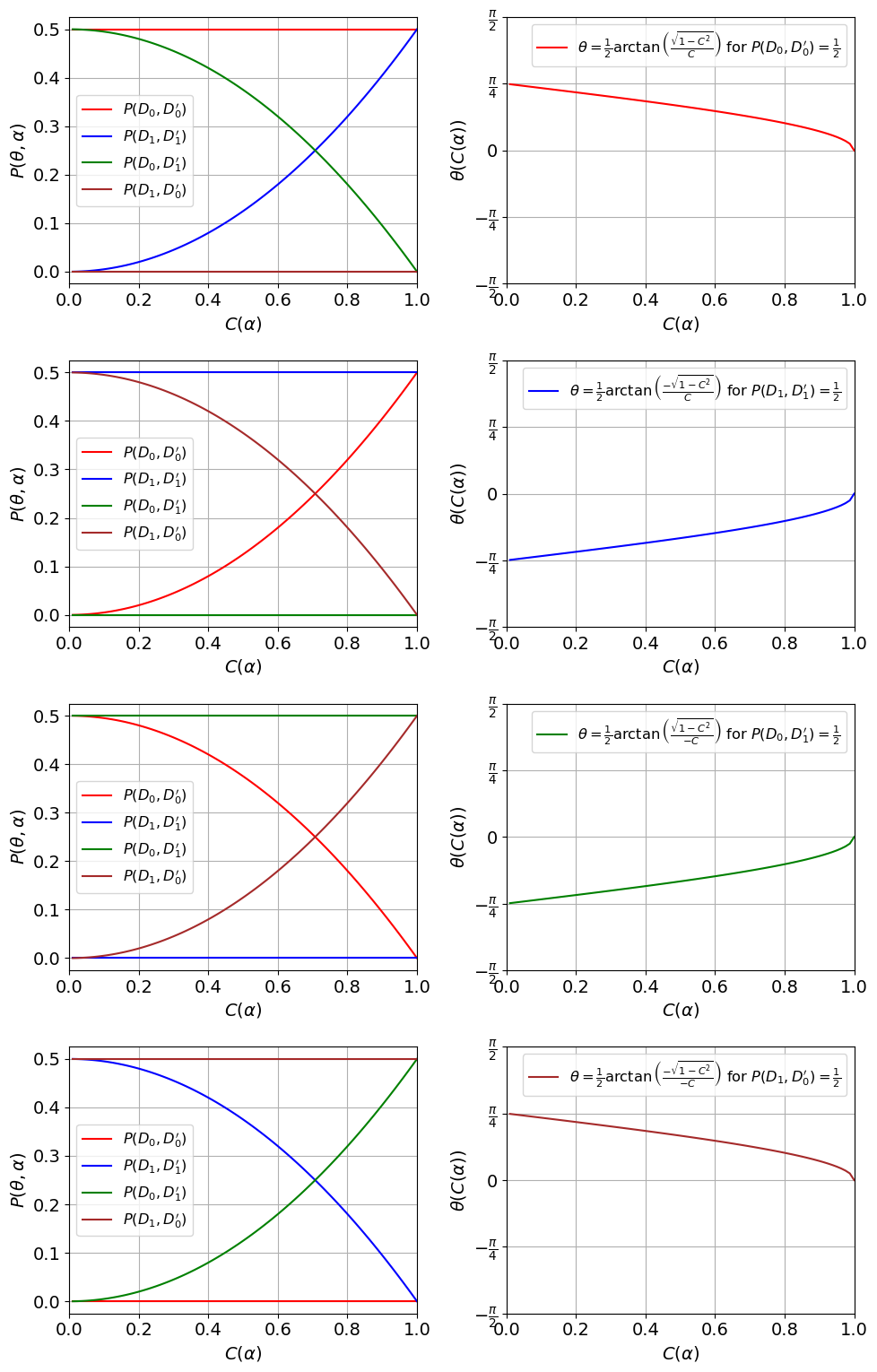}
        \caption{ In the left column, the effect of the concurrence on other probabilities is presented for angle values that fix the relevant probability at $1/2$. In the right column, it is shown how the angle that fixes the relevant probability at 1/2 changes with respect to the concurrence.}
        \label{fig:fig5}
    \end{minipage}
\end{figure}

\noindent For $C=1$ (or equivalently for $\alpha=\pi/4$) and $\theta=0$ (corresponds to the case there are no retarders), the detection probabilities are $P(D_0, D^{'}_0)=P(D_1, D^{'}_1)=\frac{1}{2}$ and $P(D_0, D^{'}_1)=P(D_1, D^{'}_0)=0$ (See Fig. \ref{fig4}). This indicates the presence of maximally entangled quantons, meaning that upon detecting one particle, the detector for the other can be predicted with certainty which is not evident from the classical perspective. The path entanglement established through momentum conservation are analogous to entanglement problems formulated through spins/polarizations 
 
\newpage

\noindent in different setups and can be used analogously. Here, the laws of quantum mechanics essentially manifest through quantum BS. For $C=1$ and the values of $\theta=(\pi/4,3\pi/4,\ldots)$, all probabilities equal to $1/4$, resulting in product states. For the other values of $\theta$, the joint-detection probabilities are reduced to $P(D_0,D^{'}_0)=P(D_1,D^{'}_1)=\frac{1}{2}\cos^2{\theta}$ and $P(D_0,D^{'}_1)=P(D_1,D^{'}_0)=\frac{1}{2}\sin^2{\theta}$. Thus, correlations can be controlled with the retarder as well as the concurrence. These results are of interest as they complete the analogy of the common spin measurement of the apparatus. The angles correspond to spin measurement probabilities in directions that are mutually aligned at the angles $\theta$ with arbitrary directions \cite{Cildiroglu2024}. 

On the other hand, for $C=0$ and the values of $\theta=(0, \pi/2, \pi, ...)$, all probabilities are equal to 
$1/4$, and the system is in the product state as expected. For the other values of $\theta$, the probabilities of the quantons being detected by the detector $D_0$ are $P(D_0, D^{'}_0)=P(D_0, D^{'}_1)=\frac{1}{4}[1+\sin{2\theta}]$, while the probabilities of detection by $D_1$ are $P(D_1, D^{'}_0)=P(D_1, D^{'}_1)=\frac{1}{4}[1-\sin{2\theta}]$, which depends on the phase of the retarder explicitly. This also indicates the presence of various correlations even in the case of $C=0$. To better understand this, it would be appropriate to observe the effects of concurrence on other probabilities at fixed angles of  that allow us to stabilize the relevant probability $P=1/2$ (see Fig. \ref{fig:fig5}). Accordingly, as an illustration, the angle that makes the probability $P(D_0, D^{'}_0)=1/2$ in the first equation of \eqref{eq:12} is $\theta=\frac{1}{2}\arctan{\bigg(\frac{\sqrt{1-C^2}}{C}\bigg)}$. Substituting the angle $\theta$ into the other equations of \eqref{eq:12}, the probabilities are obtained as $P(D_1, D^{'}_1)=\frac{1}{2}C^2$, $P(D_1, D^{'}_0)=\frac{1}{2}(1-C^2)$, and $P(D_0, D^{'}_1)=0$.
At this point, fixing the probability at 
$1/2$ reveals that $\theta$ is a function of the concurrence. These values are provided in the right column of Fig. \ref{fig:fig5}. The angles in the other graphs are $\theta=\frac{1}{2}\arctan{\bigg(\frac{-\sqrt{1-C^2}}{C}\bigg)}$, $\theta=\frac{1}{2}\arctan{\bigg(\frac{\sqrt{1-C^2}}{-C}\bigg)}$, and $\theta=\frac{1}{2}\arctan{\bigg(\frac{-\sqrt{1-C^2}}{-C}\bigg)}$,  respectively. From this perspective, it can be stated that an analogous setup is established for joint-spin/polarization measurements in arbitrary directions involving a $\pi/2$ phase shift.

\vspace{\baselineskip}

\begin{figure*}[ht]
    \centering
    \includegraphics[width=\textwidth]{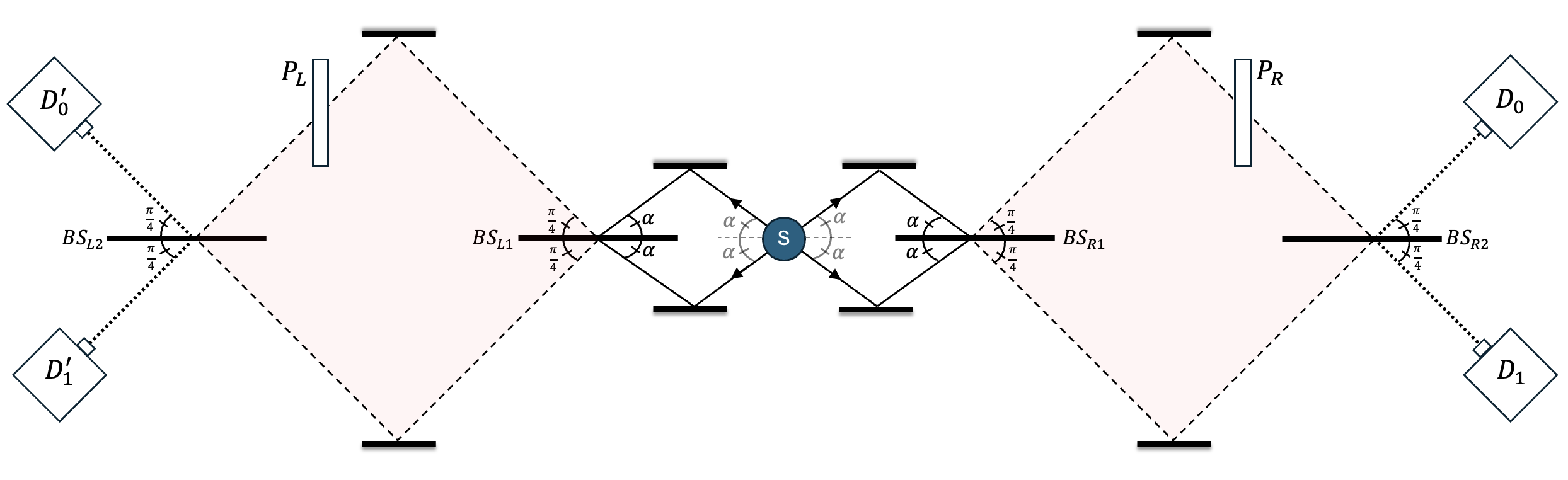}
    \caption{Schematic representation of the BS-P-BS system is presented, illustrating detection probabilities as analogs to spin measurement probabilities for spatially correlated two-quanton systems. In this setup, quantons produced in arbitrary directions, allowing the study of path entanglement and quantum correlations.}
    \label{fig6}
\end{figure*}

\noindent \textbf{\textit{BS-P-BS systems:}} In the second scenario, we generalize the system given in in Fig. \ref{fig1}.b for two quantons  produced in arbitrary directions from the same source using MZ-type interferometers modified with phase retarders (see Fig. \ref{fig6}). Thus, in the system initially described by \eqref{eq:9}, quantons pass through the first BSs and then reach the retarders placed in one of the arms. Afterward, they pass through the second BSs and are detected by the mutual detectors. In this new configuration, the final state of the system is $|\psi_{out}\rangle=(BS_{R2}\otimes BS_{L2})(P_R \otimes P_L)(BS_{R1} \otimes BS_{L1})|\psi\rangle$ and the resulting joint-detection amplitudes are, 

\begin{eqnarray}
    \label{eq:13}
    A(D_0, D^{'}_0)= \frac{\sqrt{1-C}}{2} \cos{\frac{\theta_-}{2}}+ \frac{\sqrt{1+C}}{2} \sin{\frac{\theta_+}{2}} \nonumber\\
    A(D_1, D^{'}_1)= \frac{\sqrt{1-C}}{2} \cos{\frac{\theta_-}{2}}-\frac{\sqrt{1+C}}{2} \sin{\frac{\theta_+}{2}} \nonumber\\
    A(D_0, D^{'}_1)= \frac{\sqrt{1-C}}{2} \sin{\frac{\theta_-}{2}}+\frac{\sqrt{1+C}}{2} \cos{\frac{\theta_+}{2}} \\
    A(D_1, D^{'}_0)= -\frac{\sqrt{1-C}}{2} \sin{\frac{\theta_-}{2}} +\frac{\sqrt{1+C}}{2} \cos{\frac{\theta_+}{2}} \nonumber
\end{eqnarray}

\noindent Similar to the previous section, the joint-detection probabilities can be calculated using \eqref{eq:13} amplitudes:  $P(D_i, D^{'}_j) = \left| A(D_i, D^{'}_j) \right|^2$. To simplify these expressions, wlog, let us again choose the phase of one of the retarder as $\theta_L = 0$. In this case, with $\theta_- = \theta_+ = \theta$, the probability expressions depend on both $\theta$ and $C$ are obtained, 

\begin{eqnarray}
    \label{eq:14}
    P(D_0, D^{'}_0) = \frac{1}{4}\bigg[1 - C\cos{\theta} + \sqrt{1 - C^2}\sin{\theta} \bigg]\nonumber\\
    P(D_1, D^{'}_1) = \frac{1}{4}\bigg[1 -C\cos{\theta} - \sqrt{1 - C^2} \sin{\theta} \bigg]\nonumber\\
    P(D_0, D^{'}_1) = \frac{1}{4}\bigg[ 1 + C\cos{\theta} + \sqrt{1 - C^2} \sin{\theta} \bigg] \\
    P(D_1, D^{'}_0) = \frac{1}{4}\bigg[1 + C\cos{\theta} - \sqrt{1 - C^2} \sin{\theta} \bigg] \nonumber
\end{eqnarray}

\begin{figure}[H]
    \begin{minipage}{\linewidth}
        \centering
        \includegraphics[width=\linewidth]{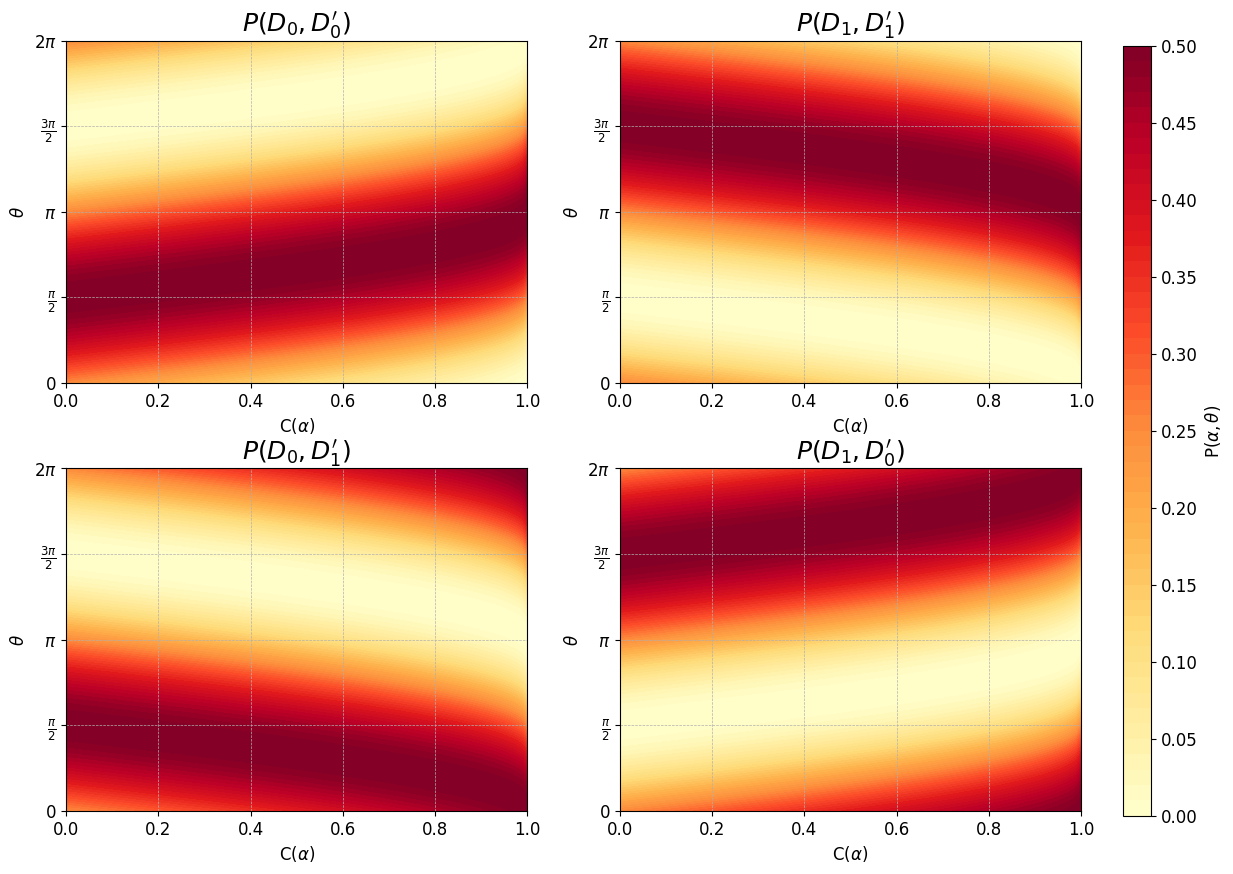}
        \caption{The probabilities in (14) depend on $\theta$ and $C(\alpha)$, shown here as a two-dimensional heat map of $P(C(\alpha), \theta)$.}
        \label{fig:Fig7}
        \includegraphics[width=\linewidth]{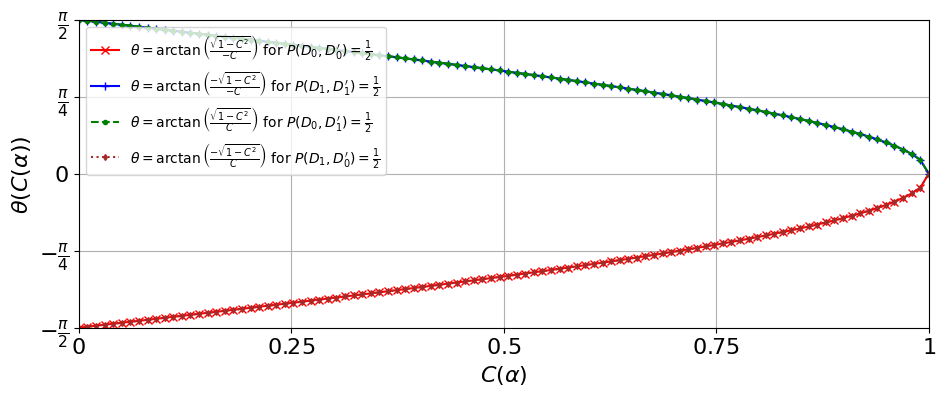}
        \caption{The phase values that stabilize the associated probabilities at 1/2 remain consistent across specific settings.}
        \label{fig:Fig9}
        \includegraphics[width=\linewidth]{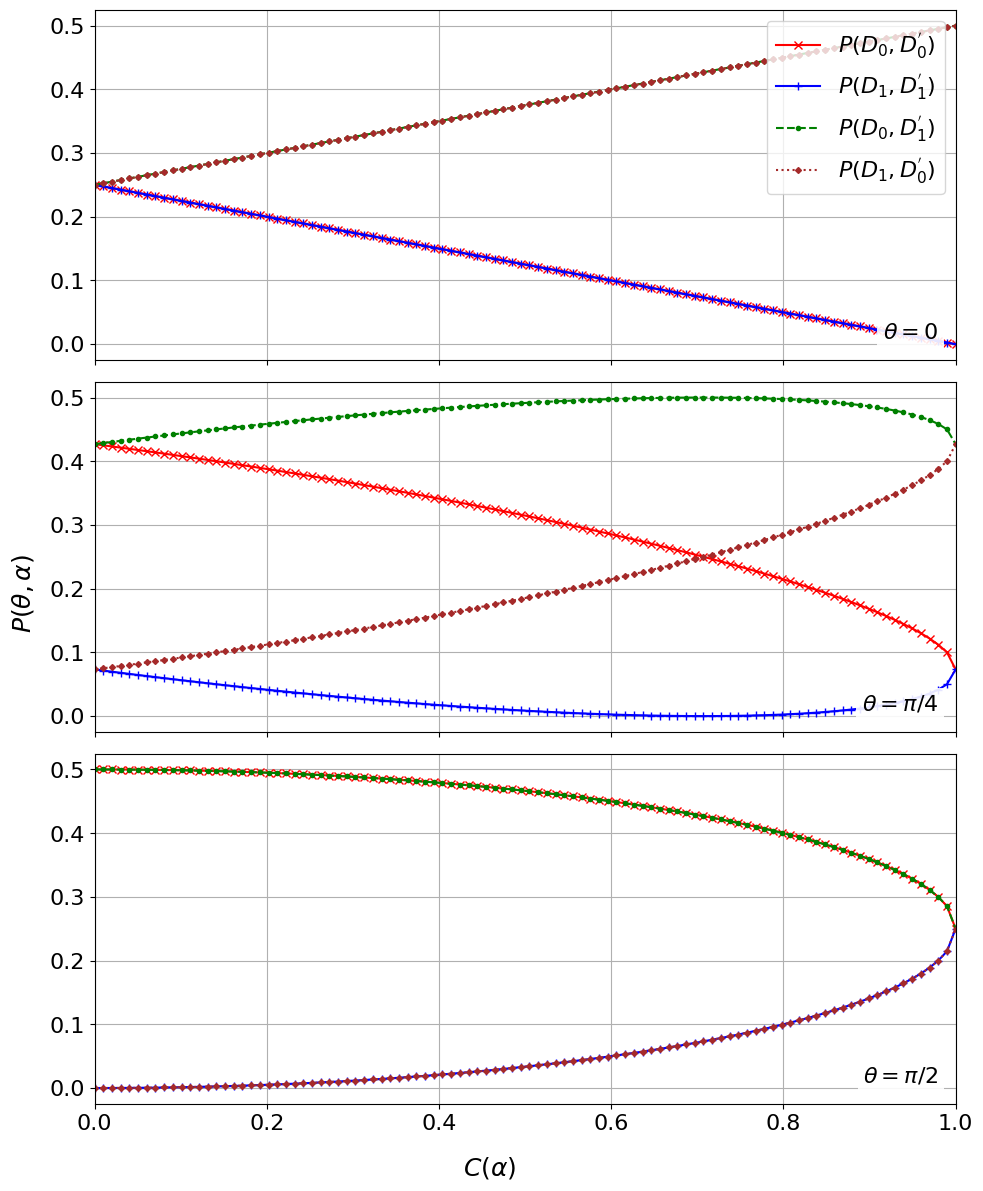}
        \caption{Joint-detection probabilities as a function of $C(\alpha)$ for fixed phases. When $C=1$ and $\theta=0$, the system is in a product state, becoming entangled at $\theta = \pi/2$. Similarly, for $C=0$, the system transitions from a product state at $\theta=\pi/2$ to a maximally entangled state at $\theta=0$. Symmetric patterns appear for retarder phases beyond $\pi/2$.}
        \label{fig:Fig8}
    \end{minipage}
\end{figure}

\noindent For $C=1$ and $\theta=0$, the expressions become $P(D_0, D^{'}_0)=P(D_1, D^{'}_1)=0$ and $P(D_0, D^{'}_1)=P(D_1, D^{'}_0)=1/2$, indicating the manifestation of maximum entanglement (See Fig. \ref{fig:Fig7}). 

\newpage

\noindent When the probabilities compared to \eqref{eq:12} (and Fig. \ref{fig4}), these results highlight a $\pi/2$ phase shift in the joint-detection probabilities \cite{Degiorgio1980}. Accordingly, it appears that the probabilities $P(D_0, D^{'}_0)=P(D_1, D^{'}_1)$ are replaced by the corresponding probabilities $P(D_0, D^{'}_1)=P(D_1, D^{'}_0)$. Hence, in this scenario the obtained results are equal to the joint-spin measurement probabilities without any approximation or phase shift. This provides a perfect analogy for the experiments conducted on spin/polarization states with spatially correlated particles. For $C=1$ and other angle values, the probabilities are reduced to $P(D_0, D^{'}_0)=P(D_1, D^{'}_1)=\frac{1}{2}\sin^2{\frac{\theta}{2}}$ and 
$P(D_0, D^{'}_1)=P(D_1, D^{'}_0)=\frac{1}{2}\cos^2{\frac{\theta}{2}}$, which are well-known probability expressions from Bell experiments \cite{Cildiroglu2024}. For $C=0$ and $\theta=0$, the system in the product state has detection probabilities where all probabilities in \eqref{eq:14} are $1/4$. For the other values of $\theta$ when $C=0$, the probabilities depend on the values of $\sin{\theta}$. In this case, for $\theta=\pi/2$, $P(D_0, D^{'}_0)=P(D_1, D^{'}_1)=\frac{1}{2}$ and $P(D_0, D^{'}_1)=P(D_1, D^{'}_0)=0$; for $\theta=3\pi/2$, the probabilities become $P(D_0, D^{'}_0)=P(D_1, D^{'}_1)=0$ and $P(D_0, D^{'}_1)=P(D_1, D^{'}_0)=\frac{1}{2}$. The phases fixing the associate probabilities at $1/2$ are given in the left column of Fig. \ref{fig:fig5} remain unchanged for the values $\theta=\frac{1}{2}\arctan{\bigg(\frac{\sqrt{1-C^2}}{-C}\bigg)}$, $\theta=\frac{1}{2}\arctan{\bigg(\frac{-\sqrt{1-C^2}}{-C}\bigg)}$, $\theta=\frac{1}{2}\arctan{\bigg(\frac{\sqrt{1-C^2}}{C}\bigg)}$, and $\theta=\frac{1}{2}\arctan{\bigg(\frac{-\sqrt{1-C^2}}{C}\bigg)}$ as presented in Fig. \ref{fig:Fig9}. Last, for fixed retarder phases $\theta=(0,\pi/4,\pi/2)$, the relationship between the concurrence and the joint-detection probabilities is particularly intriguing. Fig. \ref{fig:Fig8} illustrates how the probabilities vary with changes in $C(\alpha)$. When $C=1$ and $\theta=0$, the system is in product state; however, it becomes an entangled state when $\theta=\pi/2$. Similarly, for $C=0$ and $\theta=\pi/2$, the system initially in a product state and becomes maximally entangled at $\theta=0$. For retarder phases greater than $\pi/2$, symmetrical plots emerge, with the green-red and blue-brown lines interchanging. Ultimately, the proposed system provides substantial advantages for experimental implementation. The advantages of the MZ are revealed and the system is constructed through closed trajectories for enabling the analysis of other quantum mechanical processes that appear from the common origin.

\section{Discussion and Conclusion}

In this study, we construct a path-entangled system that exhibits correlations not classically evident, enabled uniquely by the inherently quantum-mechanical nature of the BS. One of the most intriguing uses of BS arises when paired with phase retarders in MZ-type interferometers. In such setups, the detection probabilities of quantons passing through the system can be utilized in a way that is analogous to spin/polarization measurements. Specifically, the detection probability of a quanton in a particular path corresponds to measuring its spin/polarization in an arbitrary direction. In this regard, the angle at which the spin measurement is performed can be precisely determined by the phase introduced by the retarder. By adjusting the phase, one can manipulate the detection probabilities of the quantons, which corresponds to measuring their spin in a particular direction. This connection between the retarder’s phase and spin measurements enables the interferometer setup to simulate and explore quantum spin dynamics through the behavior of quantons, thereby providing a highly flexible tool for quantum mechanical experiments. Hence, a proper basis for entanglement can be established, which allows for the testing of non-local and non-contextual properties of quantum systems.

We have explored the behavior of quanton systems generated from a source in arbitrary directions, subsequently passing through P-BS and BS-P-BS configurations. We initially analyze a system with quantons following predetermined paths, and examine the impact of variation of the production angles $\alpha$ on the final states and their detection probabilities.
We then extend the analysis to two-quanton systems, which are optimal for defining correlated quantum states. Within this framework, we establish spatial-correlations via the conservation of momentum for particles generated from the same source. We derive an expression characterizing these correlations, demonstrating that the system can be parametrized by $\alpha$. We demonstrate that the production angles correspond to the concurrence, a key entanglement measure offering a quantitative description of entanglement within the system, motivated by the fact that for certain values of $\alpha$ the system reaches different states, from product states to maximally entangled states. Thus, we have investigated the joint-detection probabilities in two possible scenarios.

In the first scenario, we considered a system where quantons are generated from a source at specific angles, and pass through phase shifters before BS. The resulting output state, expressed in terms of right and left detector bases, shows that the joint-detection amplitudes depend on specific combinations of phases of the retarders $\theta_R$ and $\theta_L$. Here, wlog, we take $\theta_L=0$ and $\theta_-=\theta_+=\theta$ to simplify the probability expressions. Calculated joint-detection probabilities reveal key behaviors: for maximal concurrence (C = 1) and no phase shift ($\theta = 0$), the system exhibits maximal entanglement, as probabilities $P(D_0, D^{'}_0)$ and $P(D_1, D^{'}_1)$ reach 1/2, while cross-detection probabilities vanish, indicating perfect correlation. Alternatively, for specific phase values ($\theta = \pi/4, 3\pi/4,\ldots$), all probabilities converge to 1/4, indicating product states. For values between these extremes, the phase $\theta$ enables control over the degree of correlation, providing an adjustable transition from product to entangled states. For the case where $C=0$, indicating no entanglement, and phase angles $\theta = 0, \pi/2, \pi, ...$, all joint-detection probabilities are uniformly equal to 1/4, resulting in a product state as expected. When $\theta$ takes other values, the system exhibits distinct probabilities for detection at different detectors. This phase dependence suggests that even when $C = 0$, certain correlations are present due to phase adjustments. To illustrate how concurrence influences probability distributions, Fig. \ref{fig:fig5} presents the effects of fixing one probability at $P = 1/2$ by selecting specific angles, which highlights the functional relationship between $\theta$ and concurrence, making $\theta$ an adjustable parameter to achieve specific correlation levels.

In the second scenario, the proposed two-quanton MZ-type setup provides an effective and versatile framework for investigating entanglement properties and joint-detection probabilities in quantum systems. By controlling C and $\theta$, the system demonstrates clear analogies with standard spin and polarization correlation experiments, enabling a detailed examination of maximally entangled states (e.g., $C=1$, $\theta=0$) as well as product states. Specifically, for $C=1$, the joint-detection probabilities reflect the well-known sinusoidal dependence seen in Bell-type experiments, indicating the presence of quantum correlations. Moreover, the stability of the detection probabilities at $P=1/2$ for fixed phase values highlights the symmetrical behaviors that emerge, even when $C=0$. This adaptable control over the concurrence-phase relationship allows the system to reproduce various states, providing a robust platform for experimentally probing entanglement effects, joint-spin measurements, and other quantum mechanical processes through closed trajectories. 

The findings underscore the MZ setup's advantages in practical applications, facilitating precise phase control and state preparation in experimental quantum physics. The analogy with spin/polarization measurements further highlights the flexibility and experimental potential of the proposed setups, offering insights into how path entanglement can be manipulated and studied in various quantum mechanical contexts through its dependency on the retarder phase and production angle, and consequently, on the concurrence. This flexibility allows for precise control of the system's entanglement, which can be exploited in various quantum mechanical applications, such as entanglement measurements and advanced quantum experiments, while the analysis present a robust framework for exploring path entanglement and the other related quantum phenomena.

\section*{Acknowledgements}

\noindent The author extends sincere gratitude to Prof. Ali Ulvi Yılmazer, Prof. Abdullah Vercin, and Dr. Melik Emirhan Tunalıoğlu for their invaluable contributions to the preliminary stages of this work. Further thanks are given to Prof. Anatoli Polkovnikov for providing the support and research opportunities that were essential in advancing this study.

\bibliographystyle{elsarticle-num}
\bibliography{refs} 

\end{document}